\documentclass[twocolumn,secnumarabic,amssymb, amsmath, nofootinbib,tightenlines,nobibnotes, aps, prl]{revtex4-1}
\usepackage{longtable}
\usepackage{bm}
\expandafter\ifx\csname package@font\endcsname\relax\else
 \expandafter\expandafter
 \expandafter\usepackage
 \expandafter\expandafter
 \expandafter{\csname package@font\endcsname}%
\fi
\usepackage{overpic}
\usepackage{tikz}
\newcommand{\beq}{\begin{equation}}
\newcommand{\eeq}{\end{equation}}
\newcommand{\lb}{\left(}
\newcommand{\rb}{\right)}
\newcommand{\un}[1]{\,\mathrm{#1}}
\usepackage{url}
\usepackage{mathtools}

\usepackage{hyperref}  

\begin{document}

\title{Gunwale bobbing}

\author{Graham P. Benham$^{a,b}$}
\email{graham.benham@ladhyx.polytechnique.fr}
\author{Olivier Devauchelle$^c$}
\author{Stephen W. Morris$^d$}
\author{Jerome A. Neufeld$^{a,b,e}$}
\affiliation{$a$: Department of Earth Sciences, University of Cambridge, Cambridge CB3 0EZ, UK}
\affiliation{$b$: Centre for Environmental and Industrial Flows, University of Cambridge, Cambridge CB3 0EZ, UK}
\affiliation{$c$: Universit\'{e} de Paris, Institut de Physique du Globe de Paris, CNRS, F-75005 Paris, France}
\affiliation{$d$: Department of Physics, University of Toronto, 60 St. George St., Toronto, ON Canada M5S 1A7}
\affiliation{$e$: Department of Applied Mathematics and Theoretical Physics, University of Cambridge, Cambridge CB3 0WA, UK}

\date{\today}%

\begin{abstract}
It has been shown experimentally that small droplets, bouncing on a vibrated liquid bath, can 
``walk" across the surface due to their interaction with their own wave-field.
Gunwale bobbing is a life-size instance of this phenomena in which a person standing on the gunwales of a canoe propels it by pumping it into oscillation with the legs. The canoe moves forward by surfing the resulting wave-field.
After an initial transient, the canoe achieves a cruising velocity which satisfies a balance between the thrust generated from pushing downwards into the surface gradients of the wave-field and the resistance due to a combination of profile drag and wave drag. By superposing the linear wave theories of Havelock (1919) for steady cruising and of Helmholtz for an oscillating source, we demonstrate that such a balance can be sustained. We calculate the optimal parameter values to achieve maximum canoe velocity. 
We compare our theoretical result to accelerometer data taken from an enthusiastic gunwale bobber. 
We discuss the similarities and differences between gunwale bobbing and hydrodynamic quantum analogues, and possible applications to competitive sports.
\end{abstract}

\maketitle


A canoe, or any small boat, can be propelled forward by forcing it into oscillation by standing on its gunwales near one end and jumping up and down.  This technique, known as gunwale bobbing, is well known to canoeists but has so far not been accounted for hydrodynamically.  We propose that the thrust sufficient to overcome drag is the result of repeatedly pushing into the surface gradients of the wave-field generated by the forced oscillation itself.  
%
In this respect, gunwale bobbing is similar to the ``walking" motion of a small drop bouncing on a vertically vibrated bath~\cite{couder2005walking}, a system that exhibits a striking analogy to Bohmian pilot wave quantum mechanics~\cite{molavcek2013drops}.  Here, we exploit this analogy by adapting recent theories of bouncing droplets~\cite{devauchelle2020walkers,andersen2015double} to treat the case of gunwale bobbing.   In addition to solving an interesting nonlinear dynamics problem that connects canoeing with quantum mechanics, an understanding of gunwale bobbing may find practical application in the general science of water sports. For example, it is not known how the rhythmic motion of the athletes' bodies during each stroke contributes to the speed or efficiency of rowing.

\begin{figure}
\centering
\begin{tikzpicture}[scale=0.6]
\node at (0,0) {\includegraphics[width=0.45\textwidth]{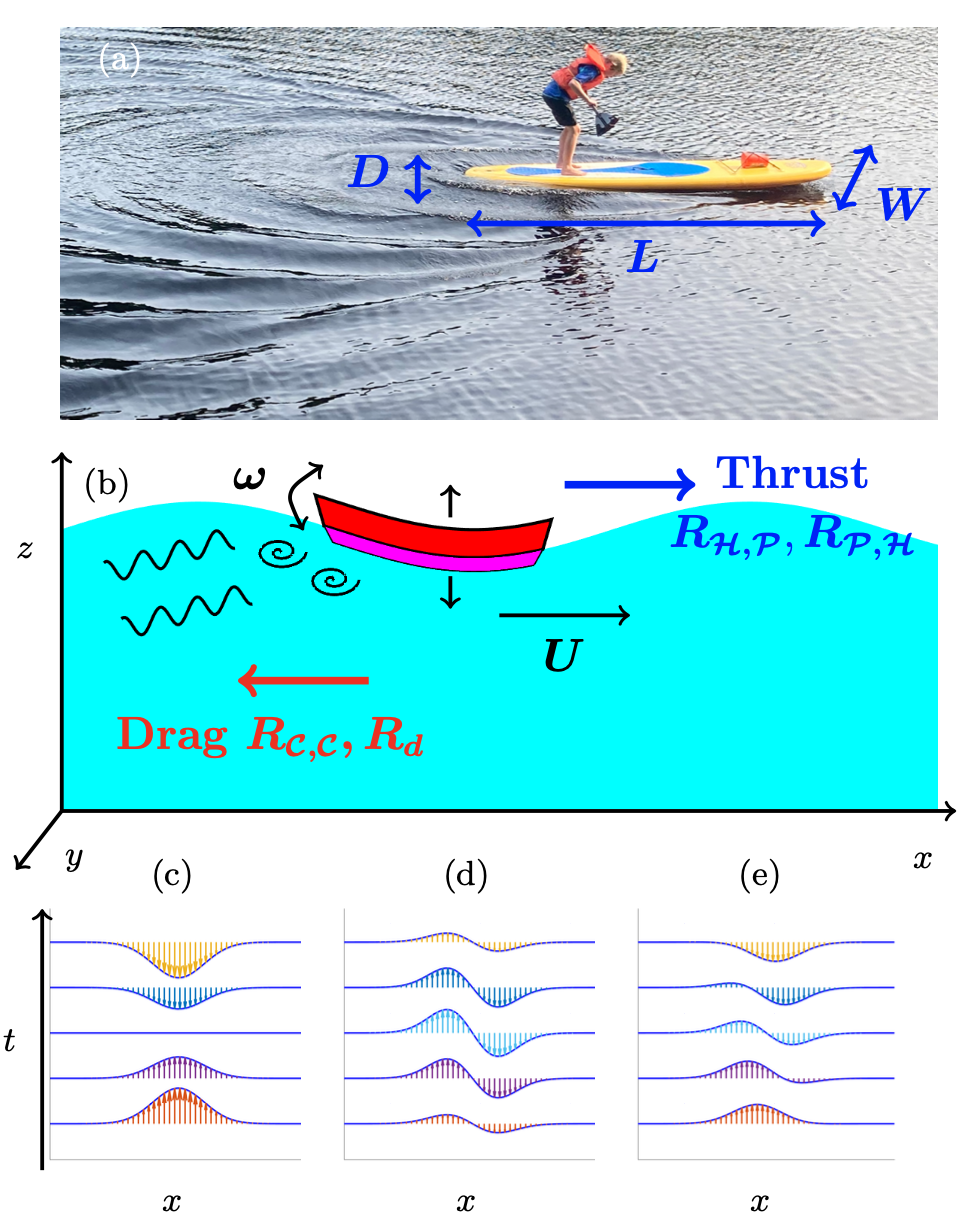}};
\end{tikzpicture}
\caption{(a) Gunwale bobbing in action, illustrating the length $L$, width $W$ and draft $D$ of the boat (here a paddleboard). (b) Illustration of the thrust and drag forces at play. Thrust originates from pushing into surface gradients producing pitching and heaving forces, $R_{\mathcal{H},\mathcal{P}},R_{\mathcal{P},\mathcal{H}}$, at frequency $\omega$, causing the boat to ``surf" on its own wave at cruising speed $U$.  Drag originates from wave energy radiation (wave drag, $R_{\mathcal{C},\mathcal{C}}$), skin friction and vortex separation (profile drag, $R_d$).  (c,d,e) Pressure source term (horizontal slice) in the case of heaving motion \eqref{heave_p}, pitching motion \eqref{pitch_p}, and heaving-pitching motion in equal proportion ($\phi=1/2$) and out of phase ($\theta=\pi/2$). \label{pitch} }
\end{figure}


In the following, we seek to optimize the forcing in order to maximize the cruising speed.  Our description is based on linear wave theory applied to a slender canoe shape moving over the water surface.  The efforts of the canoeist are assumed to produce both pitching (fore and aft) and heaving (vertical) periodic motions. We demonstrate that certain combinations of pitching and heaving produce a positive thrust force; the canoe ``surfs its own waves" (see also Ref. \cite{yuan2021wave} which discusses wake-surfing ducklings). We balance this positive thrust against negative resistive forces due to wave and profile drag to find the steady cruising speed.  For typical canoe parameters, we find that $\sim 1$~m/s cruising speeds can be achieved, in broad agreement with observations.

Consider the wave-field generated by a canoe of length $L$, width $W$ and draft $D$, moving at the surface of an infinite body of water, as illustrated in Fig. \ref{pitch}a,b, 
We take the water to be irrotational and inviscid, with density $\rho$. When gunwale bobbed, the canoe undergoes both oscillating and cruising motion. If we denote the cruising speed by $U$ and the oscillating frequency by $\omega$, then the two dimensionless parameters which define the motion are the velocity-based and frequency-based Froude numbers,
\beq
\mathrm{Fr}={U}/{\sqrt{gL}},\quad \mathrm{Fr}_\omega=\omega^{-1}\sqrt{g/L},
\eeq 
which we will use throughout this study.
We further define two aspect ratios $\alpha=L/W$ and $\beta=L/D$, both of which may be assumed to be large for slender canoes (see  Ref.~\cite{boucher2018thin} for typical parameters).
Henceforth all variables in this study are given in dimensionless form, with lengths scaled by $L$, forces scaled by $\rho U^2L^2$, pressures scaled by $\rho g L$, and time scaled by $\sqrt{L/g}$.
Following previous 
work \cite{boucher2018thin,benzaquen2011wave,benzaquen2014wake}, we assume a simplified shape for the canoe represented by a Gaussian profile
\beq
f=\exp{(-x^2/2-\alpha^2y^2/2)}.\label{gaussian}
\eeq
Our goal is to describe simultaneous heaving, pitching and cruising motion, as illustrated in Fig. \ref{pitch}c,d,e.
However, since we restrict our attention to small amplitude perturbations, we 
can treat the waves due to heaving, pitching and cruising all separately. The case of pure cruising at constant speed $U$ was addressed by Havelock \cite{havelock1919wave}. 
The motion is treated as a translating pressure source applied to the water surface
\beq
p_\mathcal{C}=  \delta \beta^{-1}f(x-\mathrm{Fr}\,t,y),\label{cruise}
\eeq
with magnitude proportional to the draft of the boat and $\delta$, an empirically determined constant. For the purposes 
of this study we will use the same value $\delta=0.4$ as Ref.~\cite{benham2020hysteretic}, which was fitted against experimental data for some small-scale boats. To 
model the pressure sources for the heaving and pitching motion, we generalize \eqref{cruise} by assuming 
\begin{align}
p_\mathcal{H}&=\mathrm{Im} \left\{ p_\mathcal{C} e^{i \mathrm{Fr}_\omega^{-1}t } \right\},~~~~~~~~~~~~~~~~~~~~~{\rm (heaving)}\label{heave_p}\\
p_\mathcal{P}&=\mathrm{Im} \left\{ p_\mathcal{C} {(x-\mathrm{Fr}\,t)} e^{i (\mathrm{Fr}_\omega^{-1} t + \theta)}  \right\},~~~{\rm (pitching)}\label{pitch_p}
\end{align}
where $\theta$ is the phase difference between heave and pitch. 
For small amplitude perturbations, it suffices to take the forcing to be a linear combination of terms,
\beq
p=p_\mathcal{C}+\phi\, p_\mathcal{P} + (1-\phi) p_\mathcal{H},\label{p_combined}
\eeq 
where $\phi\in[0,1]$ is the heave-pitch ratio. Hence, the total wave-field height resulting from this pressure disturbance \eqref{p_combined} may be split into a superposition of the corresponding wave-field components $h=h_\mathcal{C}+\phi h_\mathcal{P} + (1-\phi) h_\mathcal{H}$.

The wave drag due to the cruising disturbance \eqref{cruise} is calculated by 
noting that the pressure is applied in the manner of a rigid lid fitted to the water surface \cite{havelock1919wave,benzaquen2011wave}. 
Hence, the wave drag is given by the pressure resolved in the $x$ direction, such that the resistive force for cruising is
\beq
R_{\mathcal{C},\mathcal{C}}=\frac{1}{\mathrm{Fr}^2}\iint\limits_{-\infty}^{+\infty} p_\mathcal{C}\frac{\partial h_\mathcal{C}}{\partial X}\,\mathrm{d}X\,\mathrm{d}y,\label{cruisefor}
\eeq
where $X=x-\mathrm{Fr}\,t$ is the horizontal coordinate in the moving reference frame. 
%
%
%
%

By analogy, we may assume that forces similar to the wave 
drag  \eqref{cruisefor} arise due to the interaction between the different pressures and wave-fields due to heaving, pitching and cruising. We represent these forces, averaged over one period $T=2\pi\mathrm{Fr}_\omega$,  as
\beq
R_{n,m}=\frac{1}{\mathrm{Fr}^2 T}\int_0^T\left[\iint\limits_{-\infty}^{+\infty} p_n\frac{\partial h_m}{\partial X}\,\mathrm{d}X\,\mathrm{d}y\right]\,\mathrm{d}t,\label{bouncefor}
\eeq
where subscripts $n$ and $m$ may correspond to cruising $\mathcal{C}$, heaving $\mathcal{H}$, or pitching $\mathcal{P}$. 

We note that the dominant terms in $R_{n,m}$ typically come from interactions between out-of-phase heaving and pitching, which are of the form $\phi(1-\phi)(R_{\mathcal{H},\mathcal{P}}+R_{\mathcal{P},\mathcal{H}})$. This is because the in-phase interactions in $R_{n,n}$ time-average to small values (unless the wave-field is very asymmetric), and the interactions between oscillating and steady  terms, such as $R_{\mathcal{C},\mathcal{H}}$, time-average to zero. It is therefore intuitive that the largest forces occur for parameter values $\phi=1/2$ and $\theta=\pi/2$. Hence, for the rest of this study (unless stated otherwise), we fix these parameters values thus, and we set the aspect ratios of the canoe to those used in our experiments, $\alpha=5$, $\beta=31$.

In addition to the forces due to waves $R_{n,m}$, we must also account for the hydrodynamic forces due to skin friction and vortex separation, known together as profile drag $R_d$, which play a dominant role in opposing boat motion \cite{boucher2018thin}.
The profile drag $R_d$ is modelled using a drag coefficient $C_d$, with
\beq
R_d=-\frac{1}{2} {\mathcal{S}(\alpha,\beta)C_d(\alpha,\mathrm{Re})} ,\label{Rskin}
\eeq
where $\mathcal{S}$ is the dimensionless wetted surface area and $\mathrm{Re}$ is the Reynolds number.
Expressions for $\mathcal{S}$ (approximated by taking the canoe shape as the sum of two tetrahedrons) and $C_d$ (approximated using empirical formulae  \cite{boucher2018thin,hoerner1965practical,hadler1958coefficients}) are given in the Supplemental Material.

For small amplitude perturbations, the total horizontal force is the sum of the individual components 
\beq
\begin{split}
R_H=&\phi^2 R_{\mathcal{P},\mathcal{P}}+(1-\phi)^2 R_{\mathcal{H},\mathcal{H}}\\
&+\phi(1-\phi)(R_{\mathcal{H},\mathcal{P}}+R_{\mathcal{P},\mathcal{H}})+R_{\mathcal{C},\mathcal{C}}+R_d,\end{split}\label{sumforce}
\eeq
as is commonly 
found in other boating problems \cite{boucher2018thin,benham2019wave}.
Hence, following this approach, the cruising speed is determined by solving the nonlinear force balance
\beq
R_H(\mathrm{Fr};\mathrm{Fr}_\omega,\alpha,\beta,\phi,\theta)=0,\label{bouncesol}
\eeq
for $\mathrm{Fr}$, where the remaining parameters $\mathrm{Fr}_\omega,\alpha,\beta,\phi,\theta$ are fixed by the 
bobber and the boat.

\begin{figure}
\centering
\begin{tikzpicture}[scale=1]
\node at (0,0) {\includegraphics[width=0.48\textwidth]{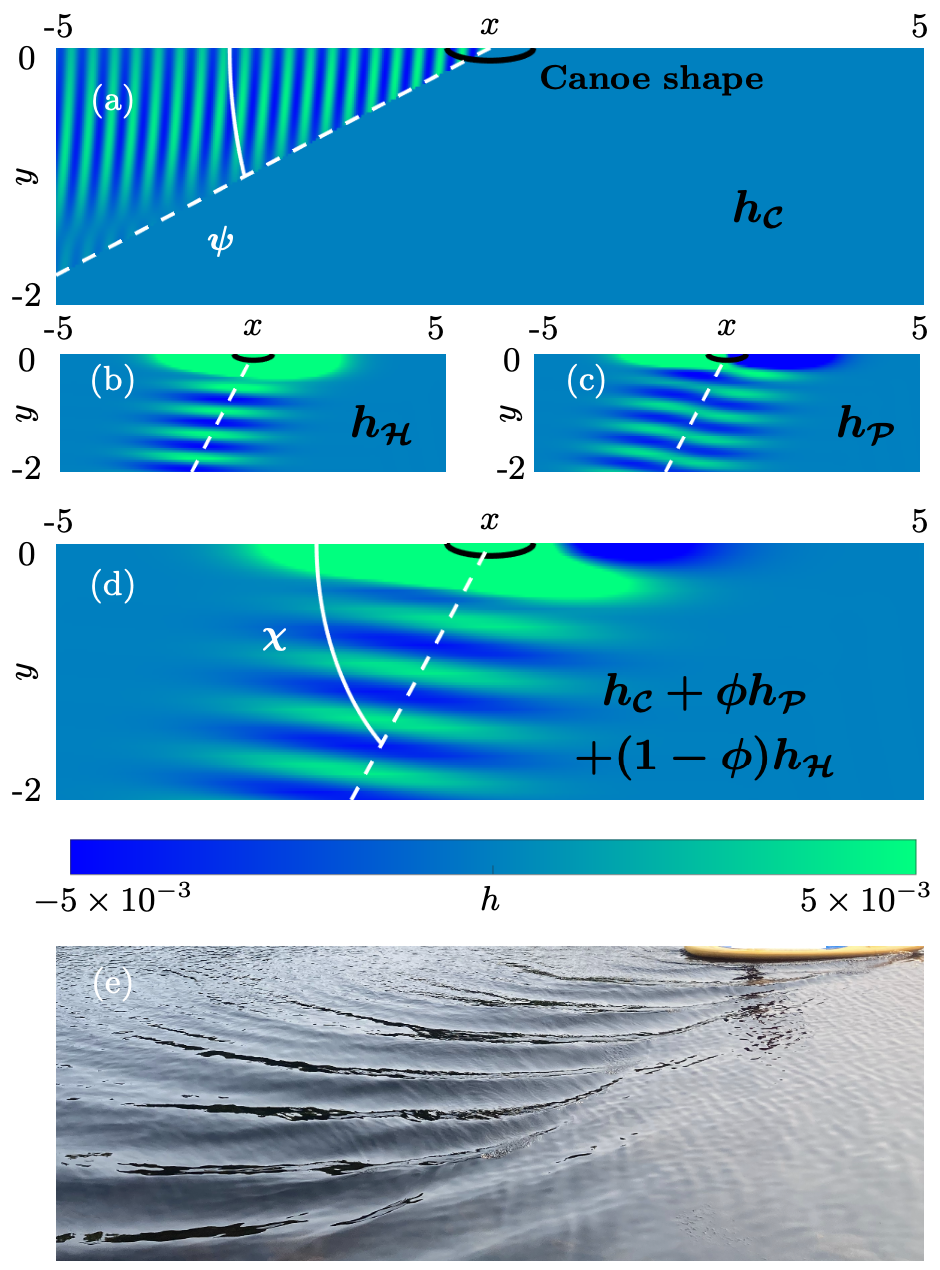}};
\end{tikzpicture}
\caption{Wave-fields 
at time $t=0$ in the case of cruising motion (a), heaving motion (b), pitching motion (c), and the combination of all 
three (d). Typical parameter values are chosen for the Froude numbers $\mathrm{Fr}=0.2$, $\mathrm{Fr}_\omega=0.25$, for which the waves due to cruising are very small, so the colour scale in (a) is $\times 10^{-5}$.
%
(e) Photo of a typical wave-field due to gunwale bobbing. \label{thewaves}}
\end{figure}

\begin{figure}
\centering
\begin{tikzpicture}[scale=1]
\node at (0,0) {\includegraphics[width=0.45\textwidth]{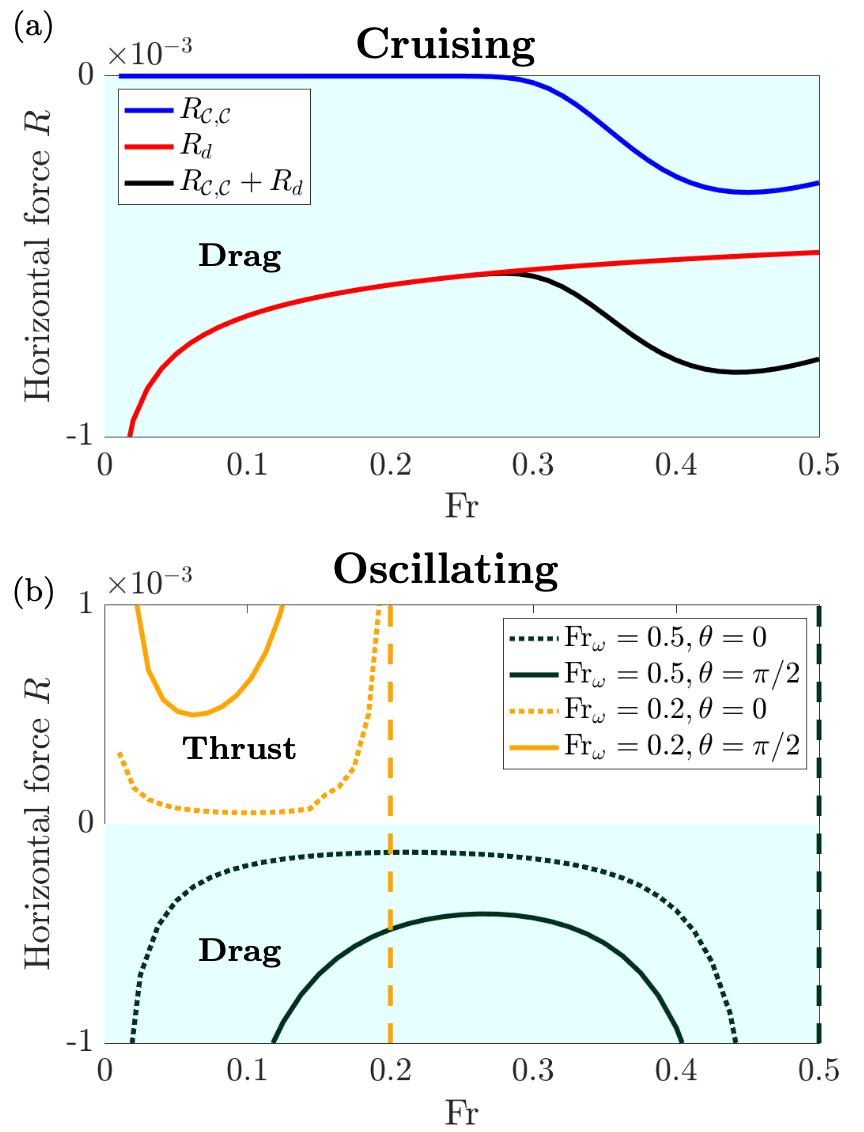}};
\end{tikzpicture}
\caption{ Horizontally resolved forces due to cruising (a) and oscillating (b). The forces due to cruising $R_{\mathcal{C},\mathcal{C}}+R_d$ are always negative (drag), whereas the summed forces due to oscillating $R_{n,m}$ (see \eqref{sumforce}) are either positive (thrust) or negative (drag) depending on the parameters $\mathrm{Fr}_\omega,\theta,\phi,\alpha,\beta$. Mach limits $\mathrm{Ma}=1$ \eqref{newFr} are indicated with dashed lines. \label{bouncethrust}}
\end{figure}

Expressions for the wave-field height $h_\mathcal{C}$ and wave resistance $R_{\mathcal{C},\mathcal{C}}$ due to steady cruising were derived by Havelock \cite{havelock1919wave} using a combination of potential flow theory and the method of Fourier transforms, assuming a small amplitude perturbation, and we display these in the Supplemental Material.  
The wave-field height $h_\mathcal{C}$ is plotted in Fig.  \ref{thewaves}a for Froude number $\mathrm{Fr}=0.2$, illustrating the famous Kelvin angle (see Refs. \cite{kelvin1887ship,rabaud2013ship,pethiyagoda2021kelvin}),
$\psi=19.47^\circ$ with dashed lines.
%
%
The cruising wave 
drag $R_{\mathcal{C},\mathcal{C}}$, which is always a negative quantity, is plotted for a variety of Froude numbers in Fig. \ref{bouncethrust}a, demonstrating the characteristic extremum at $\mathrm{Fr}\approx 0.45$. 
%
%
We compare this with the profile drag $R_d$ \eqref{Rskin} in the same plot. 
Note the change in behaviour between profile drag and wave drag that takes place near $\mathrm{Fr}\approx 0.3$. For $\mathrm{Fr}> 0.3$, a significant part of the drag comes from radiating waves away ($R_{\mathcal{C},\mathcal{C}}$), whereas for $\mathrm{Fr}< 0.3$ this becomes negligible compared to the profile drag ($R_d$).

Next, we discuss the additional waves generated by heaving and pitching motions. 
Here we follow an approach inspired by recent studies of bouncing droplets on liquid surfaces and their resultant wave-fields 
\cite{devauchelle2020walkers}. In particular, we assume that the effect of the oscillations \eqref{heave_p}-\eqref{pitch_p} is akin to the effect of a corresponding source term in a linear wave equation for the surface height $h(x,y,t)$. If the forcing frequency is $\omega$, then it is expected that the dominant component of the resultant wave-field has  a gravity-based wave-number $k=\omega^2/g$, and a gravity-based wave speed $c=g/\omega$. Hence, whilst there may be other dispersive waves in the medium, the predominant waves must travel at this speed, and hence satisfy the linear equation
\beq
\mathrm{Fr}_\omega^{-2}\frac{\partial^2 h_n}{\partial t^2}-\nabla^2 h_n = \nabla^2 p_n,\quad n=\mathcal{H},\mathcal{P}.\label{waveeq}
\eeq
The oscillating Froude number is proportional to the phase speed, $\mathrm{Fr}_\omega=c/\sqrt{gL}$, illustrating its similarity with the cruising Froude number. Hence, the Mach number for the flow is given by the ratio between these two Froude numbers
\beq
\mathrm{Ma} = \mathrm{Fr}/\mathrm{Fr}_\omega=U/c,\label{newFr}
\eeq
which must be less than one to avoid shocks.

The solution for the wave-field is calculated using a Lorentz transformation, with a Lorentz factor of $\gamma=(1-U^2/c^2)^{-1/2}$, 
and Green's functions for the Helmholtz equation \cite{devauchelle2020walkers}, the details of which can be found in the Supplemental Materials. 
The wave-field height is given by
\beq
h_n=\mathrm{Im}\left\{\mathrm{Fr}_\omega^{2}\gamma^{-1}\bar{h}_n e^{i\mathrm{Fr}_\omega^{-1}t}\right\},\quad n=\mathcal{H},\mathcal{P},\label{wavefieldsol}
\eeq
where $\bar{h}_n$ is the complex wave-field given in terms of the integrated pressure source.
The wave-field heights \eqref{wavefieldsol} for heaving, $h_\mathcal{H}$, and pitching, $h_\mathcal{P}$, are calculated numerically and plotted in Fig. \ref{thewaves}b,c, for typical parameter values.
For these 
values the canoe acts as a line source, sending waves laterally outwards, predominantly at the Mach angle (see Ref. \cite{ockendon2004waves}), $\chi=90^\circ-\tan^{-1} \mathrm{Ma}\approx 51^\circ$. The combined wave-field height due to heaving and pitching is shown in Fig. \ref{thewaves}d, showing close qualitative comparison to the waves due to gunwale bobbing in Fig. \ref{thewaves}e.

The corresponding horizontal forces due to heaving and pitching are found by numerically evaluating \eqref{bouncefor} 
(see simplifications in the Supplemental materials).
The results are plotted in Fig. \ref{bouncethrust}b for different values of the Froude number in the range $\mathrm{Fr}\in[0,\mathrm{Fr}_\omega]$, with heaving and pitching either in phase ($\theta=0$) or out of phase ($\theta=\pi/2$).
Unlike the force due to cruising, $R_{\mathcal{C},\mathcal{C}}$, which is always negative, the force due to oscillating \eqref{bouncefor} may be either negative (drag) or positive (thrust) depending on the other parameter values. Typically, the largest positive values of \eqref{bouncefor} are observed for out-of-phase pitch and heave in equal proportion, $\theta\approx\pi/2$, $\phi\approx1/2$, which appears to match with the motions of successful gunwale bobbing, as discussed later.

\begin{figure}
\centering
\begin{tikzpicture}[scale=1]
\node at (0,0) {\includegraphics[width=0.45\textwidth]{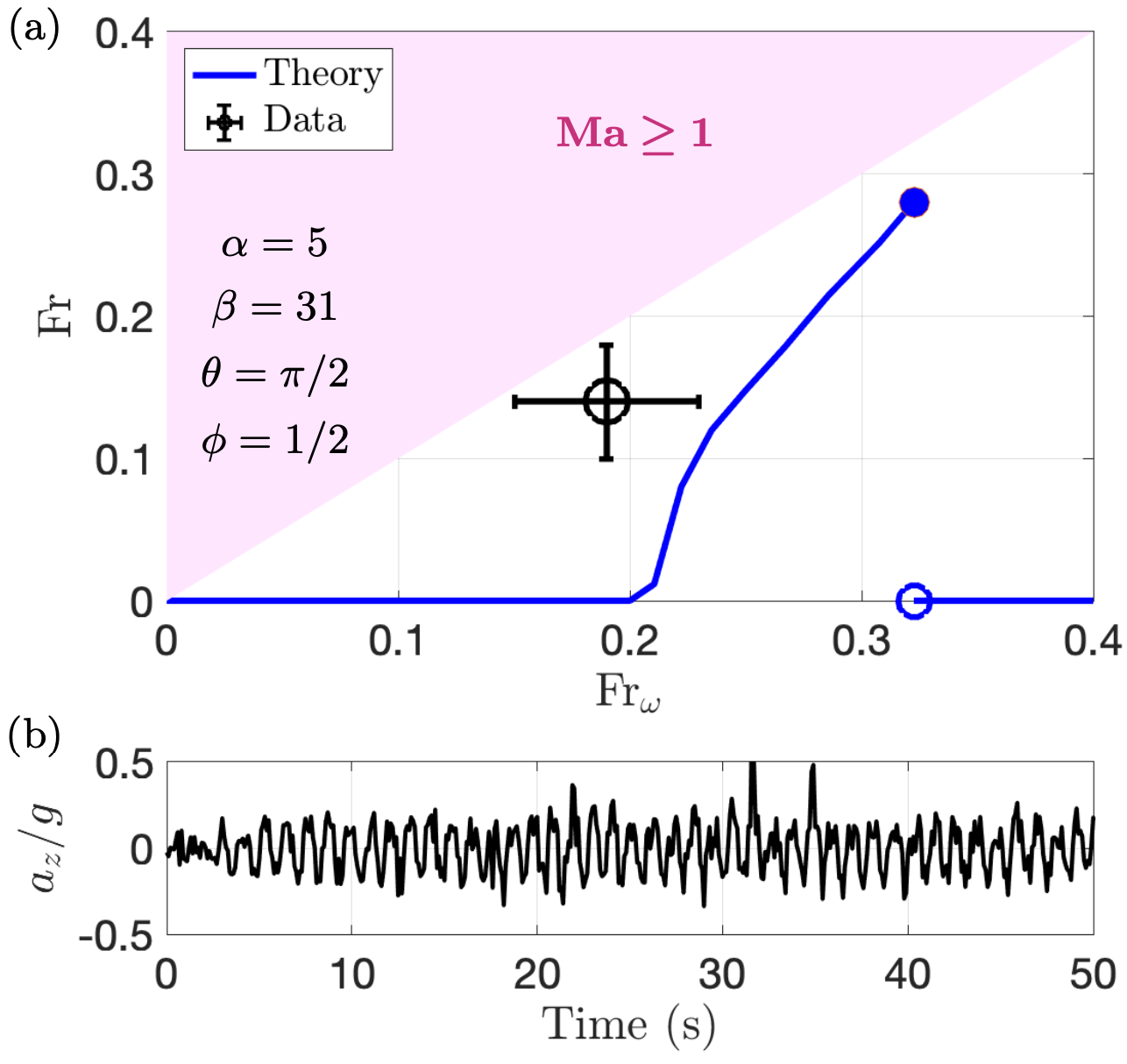}};
\end{tikzpicture}
\caption{ (a) Solutions to the thrust $-$ drag equation \eqref{bouncesol}, incorporating wave drag \eqref{cruisefor}, profile drag \eqref{Rskin}, and thrust due to oscillating \eqref{bouncefor}, for different values of the oscillating Froude number $\mathrm{Fr}_\omega$. Our experimental data point is also plotted.
Parameter space outside the Mach limit is indicated with shading. (b) A sample of the vertical acceleration divided by 
$g$, measured using an accelerometer whilst gunwale bobbing. \label{accel}}
\end{figure}

In such cases where thrust matches drag, we can solve the force balance \eqref{bouncesol} for the cruising Froude number $\mathrm{Fr}$ for gunwale bobbing. For typical parameter values there is either a single unique solution $\mathrm{Fr}$ or no solution at all (i.e. zero speed). This is plotted in Fig. \ref{accel}a for different values of the oscillating Froude number $\mathrm{Fr}_\omega$. Our results indicate that gunwale bobbing should be possible for
 the range $0.2 \leq \mathrm{Fr}_\omega \leq 0.32$ (with an abrupt cut-off when drag exceeds possible thrust values)
 and results in a cruising Froude number 
in the range $0 \leq \mathrm{Fr} \leq 0.28$.
 In this regime,  profile drag is much larger than wave drag $R_d\gg R_{\mathcal{C},\mathcal{C}}$, so 
 it provides the dominant force to be balanced against the thrust generated by oscillations.

%
%
%

We compare our theoretical results to accelerometer data taken from a gunwale bobber 
pumping one end of a canoe of length $L=4.7\un{m}$, width $W=0.94\un{m}$ and draft $D=0.15\un{m}$. The canoe was driven 
%
%
until a distance of 25$\un{m}$ was reached, taking between 20s and 50s over a series of eight trials.
The vertical acceleration $a_z$ for one of the runs is plotted in Fig. \ref{accel}b, and the remaining trials
are shown in the Supplemental Material. 
Cruising and oscillating Froude numbers were calculated as $\mathrm{Fr}= 0.14\pm0.04$ and $\mathrm{Fr}_\omega = 0.19\pm0.04$, which we illustrate with a data point in Fig. \ref{accel}a. 
It should be noted that the corresponding frequency of these oscillations is likely to have been close to the natural frequency of the canoe shape. To show this we have calculated the natural frequencies for heaving and pitching by approximating the canoe shape with two tetrahedrons, producing $\mathrm{Fr}_\omega=0.42, 0.30$, respectively (see derivation in the Supplemental Material).

It is unclear whether the forcing used in the experiments corresponds accurately with the heave and shift parameters used in our theoretical model $\phi=1/2$, $\theta=\pi/2$. 
%
%
%
However, agreement between 
experiment and theory is 
qualitatively good, though we calculate optimum Froude numbers $\mathrm{Fr}$ and $\mathrm{Fr}_\omega$ around $2\times$ and $1.7\times$ those of the experimental 
observations, respectively. It is important to note, however, 
the many simplifications used in the model such as for the shape of the canoe, the linearity of the waves, the profile drag model {\it etc.}.  
%
%
%
%
Our model nevertheless 
captures the physical mechanism behind gunwale bobbing, and provides a reasonable estimate for the permissible range of parameter values.
%

It is interesting to examine the analogy between the initial transient of gunwale bobbing and the transition to the walking state of a bouncing droplet in a bath of vibrating silicon oil, as described by Mol\'{a}\v{c}ek and Bush \cite{molavcek2013drops}. The bouncing droplets generate a circularly symmetric wave-field and only begin to walk following a symmetry-breaking bifurcation.  Our theory of gunwale bobbing assumed a fore-aft symmetric forced pitching motion; it is unknown whether a similar bifurcation might be required to start cruising.  In practice, the fore-aft symmetry is likely to be broken by the bobber standing nearer one end of the canoe. 
In general, there 
are several similarities and differences between gunwale bobbing and recently described hydrodynamic quantum analogues. Both involve thrust generated by interactions with self-generated waves, but the canoe generates traveling waves, while the bouncing droplets excite standing subcritical Faraday waves.  The two phenomena occur at vastly different Reynolds numbers.  Nevertheless, the analogy suggests some interesting possibilities for future studies. For example, one could attempt 
to have multiple canoeists 
execute dynamical waltzes, 
as in Ho {\it et al.} \cite{ho2021capillary}.

A potentially important application of  this work is to competitve water sports, and in particular rowing. During rowing races, athletes generate a significant downward force each time they generate a stroke. 
While the main propulsive thrust of a rowing boat comes from the oars, a small contribution 
may also come from the gunwale bobbing effect. 
Our present study indicates possible ways forward 
 to optimize rowing strokes to benefit from boat-wave interactions.  At Olympic-level competitions, even fractions of percentages are worth their weight in gold (medals).

\pagebreak
\widetext
\begin{center}
\textbf{\large Supplemental material for: Gunwale bobbing}
\end{center}
\setcounter{equation}{0}
\setcounter{figure}{0}
\setcounter{table}{0}
\setcounter{page}{1}
\makeatletter
\renewcommand{\theequation}{S\arabic{equation}}
\renewcommand{\thefigure}{S\arabic{figure}}
\renewcommand{\bibnumfmt}[1]{[S#1]}
\renewcommand{\citenumfont}[1]{S#1}

\section{Video link: Gunwale bobbing}

\url{https://youtu.be/XitE82v-mdY}

\section{Derivation of the wave-field for cruising}

In this section, we derive expressions for the wave-field height and wave resistance for a cruising canoe, which are based on the linear theory first developed by Havelock \cite{S_havelock1919wave}. To avoid following these derivations from first principles (since they are very long), we begin by referencing the main equations for the wave-field (in general form) and resistance taken from other literature, and continue thereon. Nevertheless, a full derivation of the theory of wave resistance can be found in \cite{S_havelock1919wave}. 

We start with the expression for the wave-field, as formulated by Benzaquen {\it et al.} \cite{S_benzaquen2014wake}, such that
\beq
h_\mathcal{C}=- \frac{1}{4\pi^2}\lim_{\epsilon\rightarrow 0} \iint\limits_{0}^{\infty}  \frac{\kappa \lb \mathcal{F} p_\mathcal{C}\rb }{\kappa-\mathrm{Fr}^2k_X^2+2i\epsilon k_X}\exp\lb{-i(k_X X+k_y y)}\rb\, \mathrm{d}k_X \,\mathrm{d} k_y,\label{hfield22}
\eeq
where $k_X,k_y$, are the wavenumbers in the $X$ and $y$ directions, $\kappa=(k_X^2+k_y^2)^{1/2}$ is the wavenumber magnitude, $\epsilon$ is a dummy asymptotic variable and $\mathcal{F} p_\mathcal{C}$ is the Fourier transform of the cruising pressure disturbance, which is
\beq
\mathcal{F} p_\mathcal{C}=\frac{\delta}{\alpha\beta} \exp\lb{-(k_X^2  +k_y^2 /\alpha^2)/4\pi^2}\rb.
\eeq
We write \eqref{hfield22} in a form which is amenable to the Sokhotski-Plemelj theorem from complex analysis, such that
\beq
h_\mathcal{C}=- \frac{\delta}{4\pi^2 \alpha \beta }\lim_{{\epsilon}\rightarrow 0} \iint\limits_{0}^{\infty}  \frac{({\kappa}/2{k}_X)  }{({\kappa}-\mathrm{Fr}^2{k}_X^2)/2{k}_X+i{\epsilon}}\exp\lb{-i({k}_X {X}+{k}_y {y})-({k}_X^2  +{k}_y^2/\alpha^2)/4\pi^2}\rb\, \mathrm{d}{k}_X \,\mathrm{d} {k}_y.\label{hfield2}
\eeq
By defining the functions 
\begin{align}
A({k}_X,{k}_y)&=({\kappa}/2{k}_X)  \exp\lb{-i({k}_X {X}+{k}_y {y})-({k}_X^2  +{k}_y^2/\alpha^2)/4\pi^2}\rb,\\
B({k}_X,{k}_y)&=({\kappa}-\mathrm{Fr}^2{k}_X^2)/2{k}_X,\label{Beqn}
\end{align}
the integral \eqref{hfield2} can be written (via a change of variables) as
\beq
h_\mathcal{C}=- \frac{\delta}{4\pi^2 \alpha \beta }\lim_{{\epsilon}\rightarrow 0} \int_0^\infty \int_{\infty}^{-\infty}  \frac{A \lb\partial B/\partial {k}_X\rb^{-1} }{B+i{\epsilon}}\, \mathrm{d}B \,\mathrm{d} {k}_y.\label{hfield3}
\eeq
Hence, upon application of the Sokhotski-Plemelj theorem, this reduces to 
\beq
h_\mathcal{C}\approx - \frac{\delta}{4\pi^2 \alpha \beta } \int_0^\infty  -i\pi A\left[{k}^*_X({k}_y),{k}_y\right]\left\{\partial B/\partial {k}_X\left[{k}^*_X({k}_y),{k}_y\right]\right\}^{-1} \,\mathrm{d} {k}_y,\label{hfield4}
\eeq
where ${k}^*_X$ is the critical wave-number for which $B=0$, which is
\beq
{k}^*_X=\mathrm{Fr}^{-2}\left[{ 1/2+\lb{ {1}/{4}+\mathrm{Fr}^4{k}_y^2}\rb^{1/2}} \right]^{1/2}.
\eeq
Note that \eqref{hfield4} ignores a second term in the Sokhotski-Plemelj theorem (proportional to the Cauchy principal value), since it is asymptotically small. The partial derivative of $B$ with respect to ${k}_X$ in \eqref{hfield4} is evaluated as \beq
\partial B/\partial {k}_X=\frac{1}{2{\kappa}}-\frac{{\kappa}}{2{k}_X^2}-\frac{\mathrm{Fr}^2}{2},
\eeq
which, evaluated at the critical wave-number is
\beq
\partial B/\partial {k}_X\left[{k}^*_X({k}_y),{k}_y\right]=\frac{1}{2\mathrm{Fr}^2\left[{k}^*_X({k}_y)\right]^2}-\mathrm{Fr}^2.
\eeq
Hence, the wave-field is ultimately written as 
\beq
h_\mathcal{C}=\frac{-i \mathrm{Fr}^{2}\delta}{4\pi\alpha\beta}\int_0^\infty {F\left[{k}_X^*({k}_y),{k}_y,{X},{y}\right]}\,\mathrm{d}{k}_y,
\eeq
where the function 
\beq
F= \frac{{\kappa}}{{k}_X(2\mathrm{Fr}^{4}-{k}_X^{-2})}\exp\lb{{-\frac{1}{4\pi^2\alpha^2}\lb{k}_X^2\alpha^{2}+{k}_y^2\rb-i\lb {k}_{X} {X}+{k}_{y} {y}\rb}}\rb.
\eeq
The corresponding wave resistance is written (following Benham {\it et al.} \cite{S_benham2020hysteretic}, except using an inertial scaling $\rho U^2L^2$ as the normalisation instead of $mg=\rho g L^3/\alpha\beta $) as
\beq
R_{\mathcal{C},\mathcal{C}}=\frac{\delta^2}{4 \mathrm{Fr}^4 \alpha^2 \beta^2} \int_0^\infty C\left[{k}^*_X({k}_y),{k}_y\right]\left\{2\mathrm{Fr}^{-2}\,\partial B/\partial {k}_X\left[{k}^*_X({k}_y),{k}_y\right]\right\}^{-1} \,\mathrm{d} {k}_y,
\eeq
where the function 
\beq
C({k}_X,{k}_y)={\kappa} \exp\lb{-({k}_X^2  +{k}_y^2/\alpha^2)/2\pi^2}\rb.
\eeq
Hence, the wave resistance simplifies to
\beq
R_{\mathcal{C},\mathcal{C}}=-\frac{\delta^2}{4 \alpha^2\beta^2}\int_0^\infty { G\left[{k}_X^*({k}_y),{k}_y\right]}\, \mathrm{d}{k}_y,
\eeq
where the function 
\begin{align}
G= \frac{{\kappa} }{2\mathrm{Fr}^{4}-{k}_X^{-2}}\exp\lb{{-\frac{1}{2\pi^2\alpha^2}({k}_X^2\alpha^2+{k}_y^2)}}\rb.
\end{align}

\section{Derivation of the wave-field for oscillating heave or pitch}  

Consider the linear wave equation
\beq
\mathrm{Fr}_\omega^{-2}\frac{\partial^2 h_n}{\partial t^2}-\nabla^2 h_n =  \nabla^2 p_n,\quad n=\mathcal{H},\mathcal{P},\label{waveeq1}
\eeq
with source terms $p_n$ given by \eqref{heave_p}-\eqref{pitch_p} in the manuscript. We apply a Lorentz transformation
\begin{align}
{X'}&=\gamma (x-\mathrm{Fr}\,t)=\gamma X,\label{lorentz1}\\
{y'}&=y,\\
{t'}&=\gamma(t-\mathrm{Fr}\mathrm{Fr}_\omega^{-2}x)=t/\gamma-\gamma \mathrm{Fr}\mathrm{Fr}_\omega^{-2} X,\label{lorentz3}
\end{align}
where $\gamma=(1-\mathrm{Fr}^2/\mathrm{Fr}_\omega^2)^{-1/2}=(1-\mathrm{Ma}^2)^{-1/2}$. 
Under the transformation \eqref{lorentz1}-\eqref{lorentz3}, the left hand side of the wave equation \eqref{waveeq1} is invariant. Hence, \eqref{waveeq1} is rewritten as
\beq
\mathrm{Fr}_\omega^{-2}\frac{\partial^2 h_n}{\partial {t'}^2}-{\nabla'}^2 h_n = \lb \gamma^2\frac{\partial^2}{\partial {X'}^2}+\frac{\partial^2}{\partial {y'}^2} \rb p_n\left[{{X'}}/{\gamma},{y'},\gamma\lb{t'}+{\mathrm{Fr}\mathrm{Fr}_\omega^{-2}{X'}}\rb\right],\label{waveeqq}
\eeq
where the original time variable $t$ is replaced using the inverse transformation identity $t=\gamma({t'}+\mathrm{Fr}\mathrm{Fr}_\omega^{-2}{X'})$. 
By inserting the expressions
\begin{align}
h_n&=\mathrm{Im}\left\{\gamma^{-1} \mathrm{Fr}_\omega^{2}h^*_ne^{i\gamma \mathrm{Fr}_\omega^{-1} t'}\right\},\label{hexpr}\\
p_n&= \mathrm{Im}\left\{ \gamma \mathrm{Fr}_\omega^{-2} p^*_ne^{i\gamma\mathrm{Fr}_\omega^{-1} t'}\right\},\label{pexpr}
\end{align}
into \eqref{waveeqq}, and by re-scaling the variables $(\tilde{X},\tilde{y})=\gamma\mathrm{Fr}_\omega^{-2} (X',y')$, we arrive at the Helmholtz equation
\beq
h^*_n+\tilde{\nabla}^2 h^*_n 
=-\gamma^2\mathrm{Fr}_\omega^{-4}\lb\gamma^2\frac{\partial^2}{\partial \tilde{X}^2}+\frac{\partial^2}{\partial \tilde{y}^2}\rb p^*_n,\label{helm}
\eeq
where the complex source terms are
\begin{align}
p^*_\mathcal{H}&=\frac{\delta\mathrm{Fr}_\omega^{2}}{\beta \gamma} \exp\lb{-{\mathrm{Fr}_\omega^{4}(\tilde{X}^2+(\alpha\gamma)^2\tilde{y}^2)}/{2\gamma^{4}}+i \mathrm{Ma} \tilde{X}}\rb,\\
p^*_\mathcal{P}&=\frac{\delta\mathrm{Fr}_\omega^{4}}{\beta \gamma^3} {\tilde{X}}\exp\lb{i\theta-{\mathrm{Fr}_\omega^{4}(\tilde{X}^2+(\alpha\gamma)^2\tilde{y}^2)}/{2\gamma^{4}}+i \mathrm{Ma} \tilde{X}}\rb.
\end{align}
As described by Devauchelle {\it et al.} \cite{S_devauchelle2020walkers}, the solution to \eqref{helm} is
\beq
h^*_n=\frac{i}{4}\iint\limits_{-\infty}^{+\infty} H_0^{(1)}(|\tilde{\mathbf{X}}-\tilde{\boldsymbol{\mathcal{X}}}|) \gamma^2\mathrm{Fr}_\omega^{-4}\lb\gamma^2\frac{\partial^2}{\partial \tilde{\mathcal{X}}^2}+\frac{\partial^2}{\partial \tilde{\mathcal{Y}}^2}\rb p^*_n(\tilde{\boldsymbol{\mathcal{X}}})\,\mathrm{d}\tilde{\boldsymbol{\mathcal{X}}},
\eeq
written in terms of the Hankel function of the first kind and zeroth order $H_0^{(1)}$ (which is the Green's function for the Helmholtz equation) and integrated over dummy variables $\tilde{\boldsymbol{\mathcal{X}}}=(\tilde{\mathcal{X}},\tilde{\mathcal{Y}})$. 
To acquire the final solution for the wave-field, we re-write expressions \eqref{hexpr} and \eqref{pexpr} in terms of the original time variable $t$, noting that
\beq
i\mathrm{Fr}_\omega^{-1} \gamma t'=i\mathrm{Fr}_\omega^{-1}  t-i\tilde{X}\mathrm{Fr}\mathrm{Fr}_\omega^{-1}=i\mathrm{Fr}_\omega^{-1}  t-i\mathrm{Ma}\tilde{X}.
\eeq
Hence, the wave-field and pressure source are given by
\begin{align}
h_n&=\mathrm{Im}\left\{\gamma^{-1}\mathrm{Fr}_\omega^{2}\bar{h}_n(\tilde{X},\tilde{y})\,e^{i\mathrm{Fr}_\omega^{-1} t}\right\},\\
p_n&=\mathrm{Im}\left\{\gamma \mathrm{Fr}_\omega^{-2}\bar{p}_n(\tilde{X},\tilde{y})\, e^{i\mathrm{Fr}_\omega^{-1} t}\right\},
\end{align}
where 
\begin{align}
\bar{h}_n&=h^*_n e^{-i \mathrm{Ma} \tilde{X}}=\left[\frac{i}{4}\iint\limits_{-\infty}^{+\infty}  H_0^{(1)}(|\tilde{\mathbf{X}}-\tilde{\boldsymbol{\mathcal{X}}}|)\mathcal{L}\bar{p}_n(\tilde{\boldsymbol{\mathcal{X}}}) e^{i\mathrm{Ma}\tilde{{\mathcal{X}}}} \,\mathrm{d}\tilde{\boldsymbol{\mathcal{X}}}\right]e^{-i\mathrm{Ma}\tilde{X}},\\
\bar{p}_n&=p^*_n e^{-i \mathrm{Ma} \tilde{X}}=\begin{cases}
\frac{\delta\mathrm{Fr}_\omega^{2}}{\beta \gamma} \exp\lb{-{\mathrm{Fr}_\omega^{4}(\tilde{X}^2+(\alpha\gamma)^2\tilde{y}^2)}/{2\gamma^{4}}}\rb,\quad & n=\mathcal{H},\\
& \\
\frac{\delta\mathrm{Fr}_\omega^{4}}{\beta \gamma^3} {\tilde{X}}\exp\lb{i\theta-{\mathrm{Fr}_\omega^{4}(\tilde{X}^2+(\alpha\gamma)^2\tilde{y}^2)}/{2\gamma^{4}}}\rb,\quad & n=\mathcal{P},
\end{cases}
\end{align}
where the operator $\mathcal{L}=\gamma^2\mathrm{Fr}_\omega^{-4}(\gamma^2\partial^2/\partial \tilde{\mathcal{X}}^2+\partial^2/\partial \tilde{\mathcal{Y}}^2)$.
Now to calculate the force $R_{n,m}$, we start with the expression for the horizontally resolved pressure force
\beq
R_{n,m}=\frac{1}{\mathrm{Fr}^2T}\int_0^T\left[\iint\limits_{-\infty}^{+\infty} p_n\frac{\partial h_m}{\partial X}\,\mathrm{d}X\,\mathrm{d}y\right]\,\mathrm{d}t.\label{bouncefor2}
\eeq
Noting that 
\beq
\frac{\partial h_n}{\partial X}=\mathrm{Im}\left\{\gamma \frac{\partial \bar{h}_n}{\partial \tilde{X}} \,e^{i\mathrm{Fr}_\omega^{-1} t}\right\},
\eeq
the force \eqref{bouncefor2} is given in dimensionless terms as
\beq
R_{n,m}=
\frac{\mathrm{Fr}_\omega^{2}}{\mathrm{Fr}^2\gamma T}\int_0^T\iint\limits_{-\infty}^{+\infty}\mathrm{Im}\left\{ \bar{p}_ne^{i\mathrm{Fr}_\omega^{-1} t}\right\} \mathrm{Im}\left\{ \frac{\partial \bar{h}_m}{\partial \tilde{X}}e^{i\mathrm{Fr}_\omega^{-1} t}\right\} \,\mathrm{d}\tilde{X}\,\mathrm{d}\tilde{y}\,\mathrm{d}t.
\eeq
However, by noticing that the time average of any product of the form
\beq
\frac{1}{T}\int_0^T \mathrm{Im}\left\{ A e^{i\mathrm{Fr}_\omega^{-1} t}\right\} \mathrm{Im}\left\{ B e^{i\mathrm{Fr}_\omega^{-1} t}\right\}   \,\mathrm{d}t=\frac{1}{2T}\int_0^T \mathrm{Im}\left\{ A\right\} \mathrm{Im}\left\{ B \right\} + \mathrm{Re}\left\{ A\right\} \mathrm{Re}\left\{ B \right\}   \,\mathrm{d}t,
\eeq
then we arrive at the final expression
\beq
R_{n,m}=\frac{\mathrm{Fr}_\omega^2}{2\mathrm{Fr}^2\gamma}\iint \limits_{-\infty}^{+\infty}\mathrm{Re} \,\bar{p}_n \,\mathrm{Re} \frac{\partial \bar{h}_m}{\partial \tilde{X}}+\mathrm{Im}\, \bar{p}_n \,\mathrm{Im} \frac{\partial \bar{h}_m}{\partial \tilde{X}} \,\mathrm{d}\tilde{X}\,\mathrm{d}\tilde{y}.
\eeq

\section{Canoe surface area and natural frequencies}

\begin{figure}
\centering
\begin{tikzpicture}[scale=1.55]
\node at (7.3,-8.5) {\includegraphics[width=0.5\textwidth]{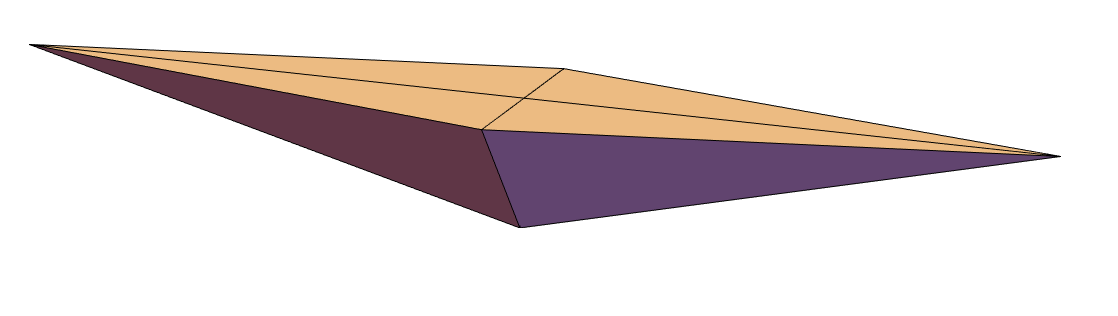}};
\node[blue] at (9.3,-8.2) {\large$\boldsymbol{L}$};
\node[blue] at (7.5,-7.8) {\large$\boldsymbol{W}$};
\node[blue] at (3.9,-8.3) {\large$\boldsymbol{D}$};
\draw[line width=1.5,<->,blue] (4.55,-7.9) -- (9.85,-8.5);
\draw[line width=1.5,<->,blue] (4.25,-8.7) -- (4.25,-7.95);
\draw[line width=1.5,<->,blue] (7.35,-8) -- (6.9,-8.35);
\end{tikzpicture}
\caption{Canoe shape approximated by two tetrahedrons with total length, width and draft, $L$, $W$ and $D$.\label{canoe}}
\end{figure}

For the sake of simplicity we approximate the surface area of the canoe by considering two tetrahedrons stuck together with the same total dimensions, as illustrated in Fig. \ref{canoe}. The wetted area of one of the exposed triangles of the tetrahedron surface can be calculated through trigonometry, giving
\beq
A=\frac{1}{8}\left[ L^2 W^2 + 4 D^2 (L^2 + W^2) \right]^{1/2}.
\eeq
Hence, in dimensionless coordinates, the total surface area (four triangles) is 
\beq
\mathcal{S}=\left[\frac{1}{4\alpha^2}+ \frac{1}{\beta^2} \lb 1 + \frac{1}{\alpha^2}\rb \right]^{1/2}.
\eeq

Next, we use this approximate canoe shape to estimate the natural frequencies for heaving and pitching. This is done by considering the simple harmonic motion of small perturbations to the vertical position of the centre of mass and pitch angle. 
To calculate these natural frequencies it is assumed that the canoe is not cruising, such that $\mathrm{Fr}=0$, $U=0$. 
Whilst all of the analysis in this study has so far remained dimensionless, we keep the subsequent analysis dimensional for convenience. To make this clear we have capitalised many variables so as not to be confused with their dimensionless counterparts in the manuscript. The two variables for which this doesn't apply are $x$ and $t$, whose capitals already have prescribed definitions in the manuscript, so we replace these with hatted capitals, $\hat{X},\hat{T}$, for clarity.  

Let the vertical position of the centre of mass (measured from the waterline) and the pitch angle of the canoe be denoted by $Z(\hat{T})$ and $\varphi(\hat{T})$ respectively. The equations of motion for each of these are given by conservation of linear and angular momentum in the heave and pitch directions, such that
\begin{align}
(m+m_a)\ddot{Z}&=-\iint_S (P-P_a) (\hat{\mathbf{n}}\cdot \hat{\boldsymbol{k}}) \,\mathrm{d}S-m g,\label{vertf}\\
(I+I_a)\ddot{\varphi}&=\iint_S (P-P_a) ( \mathbf{r}\times\hat{\mathbf{n}})\cdot \hat{\boldsymbol{\jmath}} \, \mathrm{d}S,\label{angf}
\end{align}
where $P_a$ is atmospheric pressure, $\mathbf{r}$ is the position vector measured from the origin, $\hat{\mathbf{n}}$ is the outward pointing unit normal vector to the hull surface (where $S=L^2\mathcal{S}$), $m$ is the mass of the canoe, $m_a$ and $I_a$ are the linear and angular added masses in the heave and pitch directions \cite{S_ursell1949heaving}, and $I$ is the moment of inertia which is defined as
\beq
I=\rho_0\iiint_V |\mathbf{r}|^2 \,\mathrm{d}V,\label{momi}
\eeq
in terms of the hull material density $\rho_0$ (assumed constant), and volume $V$.
For simplicity we have ignored the presence of a gunwale bobber in \eqref{momi} and instead we have assumed that the mass of the canoe is distributed uniformly over its hull.

We first consider the case of pure heaving ($\varphi=0,Z=Z(\hat{T})$) and attempt to find an approximate expression for the natural frequency. In this case, the centre of mass is perturbed vertically according to
\beq
Z=Z_0+\zeta(\hat{T}),
\eeq
where $Z_0$ is a constant and $|\zeta/Z_0|\ll 1$. Likewise, if $Z=-\Gamma(\hat{X},Y)$ is the shape function of the canoe composed of two tetrahedrons, then this is also perturbed by $Z=-\Gamma(\hat{X},Y)+ \zeta(\hat{T})$.
Since the oscillations are of small amplitude, the pressure on the hull surface is approximately hydrostatic
\beq
P\approx P_a-\rho g (-\Gamma+ \zeta) .
\eeq
Hence, the leading order and first order terms of \eqref{vertf} reduce to
\begin{align}
0&=-\rho g \iint_S (\hat{\mathbf{n}}\cdot \hat{\boldsymbol{k}}) \Gamma \,\mathrm{d}S-mg,\label{masseq}\\
(m+m_a)\ddot{\zeta}&=\left[\rho g\iint_S    (\hat{\mathbf{n}}\cdot \hat{\boldsymbol{k}})  \, \mathrm{d}S\right] \zeta,
\end{align}
the first of which sets the mass of the boat and the second of which determines the perturbation dynamics. These are neatly rearranged to give a simple harmonic oscillator equation for the small perturbation, such that
\beq
\ddot{\zeta}+ \omega^2 \zeta=0,
\eeq
where the natural frequency is 
\beq
\omega^2=\frac{ g\iint_S  (\hat{\mathbf{n}}\cdot \hat{\boldsymbol{k}})\,\mathrm{d}S}{  \iint_S (\hat{\mathbf{n}}\cdot \hat{\boldsymbol{k}}) \Gamma \,\mathrm{d}S(1+m_a/m)}.\label{nat1}
\eeq
The added mass ratio $m_a/m$ for the tetrahedron shape is unknown, so instead (as a very approximate estimate) we use the value for a heaving ellipse with the same length-depth aspect ratio (see Ref. \cite{S_newman2018marine}), which is
\beq
\frac{m_a}{m}=\frac{\pi\rho L^2/4}{\pi \rho LD/2}=\frac{\beta}{2}.
\eeq
Due to the symmetries of the tetrahedra, it suffices to consider just one of the outer triangular faces to calculate \eqref{nat1}. Taking $\Gamma$ as the planar outer face
\beq
\Gamma=D\left[1+\frac{2\hat{X}}{L}+\frac{2Y}{W}\right],\label{quarter}
\eeq
which is defined in the range $Y\in[-W(1/2 + \hat{X}/L),0]$, $\hat{X}\in[-L/2,0]$, we calculate 
\beq
\hat{\mathbf{n}}\cdot \hat{\boldsymbol{k}}=-\left[1+\frac{4}{\beta^2}(1+\alpha^2)\right]^{-1/2}=-\frac{1}{2\alpha \mathcal{S}(\alpha,\beta)},
\eeq
which is a constant. Now the natural heaving frequency $\omega$ may be calculated by evaluating the integrals in \eqref{nat1}. After simplification and re-writing in terms of the oscillating Froude number, this provides the result
\beq
\mathrm{Fr}_\omega=\frac{1}{\omega}\lb\frac{g}{L}\rb^{1/2}=\lb\frac{1+\beta/2}{3\beta}\rb^{1/2}.\label{heavenat}
\eeq

Next we repeat the above analysis for the case of pure pitching ($\zeta=0,\varphi=\varphi(\hat{T}) $), centred about the origin. A small perturbation $\varphi\ll1$ is applied to the pitch angle of the boat, causing the vertical position of the hull to be moved to $Z\approx -\Gamma+\varphi \hat{X}$. 
To evaluate the right hand side of \eqref{angf}, we need to calculate the cross product of the radial vector with the pressure force vector. Each of these vectors is given by
\begin{align}
\mathbf{r}&=\left\{\hat{X},Y,-\Gamma+\varphi \hat{X}\right\},\\
(P-P_a)\hat{\mathbf{n}}&=\frac{-\rho g (-\Gamma+\varphi \hat{X})}{\left[ \lb -\frac{\partial \Gamma}{\partial \hat{X}}+\varphi\rb^2+\frac{\partial \Gamma}{\partial Y}^2+1 \right]^{1/2}}\left\{ -\frac{\partial \Gamma}{\partial \hat{X}}+\varphi,-\frac{\partial \Gamma}{\partial Y},-1\right\}.
\end{align}
Next we perform the cross product, expand out the variables in powers of $\varphi$, and integrate over the surface area of the hull (note all anti-symmetric terms vanish upon integration), ignoring terms of order $\mathcal{O}(\varphi^2)$. This gives the result
\beq
\begin{split}
&\iint_S \lb P-P_a\rb \lb \mathbf{r}\times \hat{\mathbf{n}}\rb\cdot \hat{\boldsymbol{\jmath}} \, \mathrm{d}S\approx \\
&-\left[\rho g \iint_S  \frac{ \hat{X}^2\lb 1+\frac{\partial \Gamma}{\partial \hat{X}}^2+\frac{\partial \Gamma}{\partial Y}^2\rb+\Gamma^2\lb 1+\frac{\partial \Gamma}{\partial Y}^2\rb+\hat{X}\Gamma\frac{\partial \Gamma}{\partial \hat{X}}\lb 1+2\frac{\partial \Gamma}{\partial \hat{X}}^2+2\frac{\partial \Gamma}{\partial Y}^2\rb}{\left[ \frac{\partial \Gamma}{\partial \hat{X}}^2+\frac{\partial \Gamma}{\partial Y}^2+1 \right]^{3/2}}   \, \mathrm{d}S\right] \varphi .\label{angf2}
\end{split}
\eeq
The next step is to calculate the moment of inertia in \eqref{angf}. To do so we assume that the hull mass is distributed over a thin solid shell of vertical thickness $H\ll \Gamma$, such that the volume element is approximately $dV\approx H dS$ and the hull density $\rho_0 ={m}/{H S}$. Hence, the moment of inertia \eqref{momi} is given by
\beq
I\approx \frac{m}{S}\iint_S (\hat{X}^2+Y^2+\Gamma^2) \,\mathrm{d}S.\label{momi1}
\eeq
Note that only leading order terms are kept in \eqref{momi1} since the left hand side of \eqref{angf} is already of order $\mathcal{O}(\varphi)$.
Using \eqref{masseq} to replace $m$ (at leading order) with an expression for the mass, then \eqref{momi1} becomes
\beq
I\approx\frac{\rho}{S}\iint_S (\hat{X}^2+Y^2+\Gamma^2) \,\mathrm{d}S\iint_S -(\hat{\mathbf{n}}\cdot \hat{\boldsymbol{k}}) \Gamma \,\mathrm{d}S.
\eeq
Hence, the equation of motion \eqref{angf} reduces to a simple harmonic oscillator of the form
\beq
\ddot{\varphi}+ \omega^2\varphi = 0,
\eeq
where
\beq
\begin{split}
\omega^2=\frac{gS \iint_S  { \left[\hat{X}^2\lb 1+\frac{\partial \Gamma}{\partial \hat{X}}^2+\frac{\partial \Gamma}{\partial Y}^2\rb+\Gamma^2\lb 1+\frac{\partial \Gamma}{\partial Y}^2\rb + \hat{X}\Gamma\frac{\partial \Gamma}{\partial \hat{X}}\lb 1+2\frac{\partial \Gamma}{\partial \hat{X}}^2+2\frac{\partial \Gamma}{\partial Y}^2\rb\right]}{\left[ \frac{\partial \Gamma}{\partial \hat{X}}^2+\frac{\partial \Gamma}{\partial Y}^2+1 \right]^{-3/2}}   \, \mathrm{d}S}{
{\iint_S (\hat{X}^2+Y^2+\Gamma^2) \,\mathrm{d}S\iint_S -(\hat{\mathbf{n}}\cdot \hat{\boldsymbol{k}}) \Gamma\,\mathrm{d}S}(1+I_a/I)}.\label{compint}
\end{split}
\eeq
Since the angular added mass ratio $I_a/I$ is unknown for the tetrahedron shape, we use the value for an ellipse with the same length-depth aspect ratio as a very approximate estimate, which is 
\beq
\frac{I_a}{I}=\frac{1/8\pi\rho (L^2/4-D^2)^2}{1/8\pi \rho LD (L^2/4+D^2)}=\frac{ (\beta^2-4)^2}{ 4\beta (\beta^2+4)}.  
\eeq

As before, we exploit the symmetry of the tetrahedra to evaluate the integrals in \eqref{compint} by considering only one of the outer triangular faces of the surface \eqref{quarter}. In this way the the natural frequency for pitching may be calculated analytically, and we write this (after simplification) in terms of the oscillating Froude number, which is
\beq
\mathrm{Fr}_\omega=\left[\frac{
 \lb 4 + 4 \alpha^2 +  \beta^2\rb^{1/2} \lb \beta^2 + \alpha^2 \lb 4 + \beta^2\rb\rb}{3 \alpha^2 \lb -16 + \lb 6 + 4 \alpha^2\rb \beta^2 + \beta^4\rb}\lb 1+ \frac{ (\beta^2-4)^2}{ 4\beta (\beta^2+4)}\rb \right]^{1/2}.\label{pitchnat}
\eeq
Hence, inserting $\alpha=5$, $\beta=31$, into the (very approximate) expressions for the natural oscillating Froude number \eqref{heavenat},\eqref{pitchnat}, we calculate $\mathrm{Fr}_\omega=0.42$ in the case of heaving and $\mathrm{Fr}_\omega=0.30$ in the case of pitching.

\section{Accelerometer data}

\begin{figure}
\centering
\begin{tikzpicture}[scale=1]
\node at (0,0) {\includegraphics[width=0.2\textwidth,trim={10cm 0 0 0},clip]{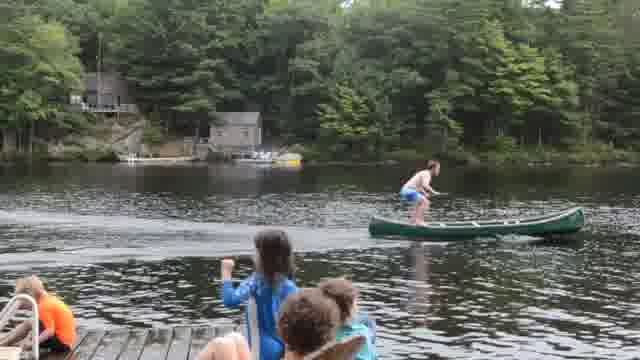}};
\node at (4,0) {\includegraphics[width=0.2\textwidth,trim={10cm 0 0 0},clip]{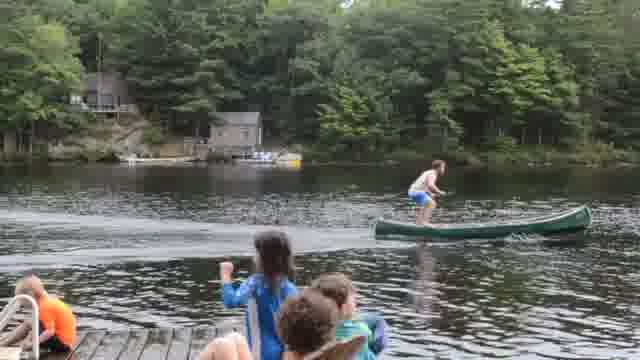}};
\node at (8,0) {\includegraphics[width=0.2\textwidth,trim={10cm 0 0 0},clip]{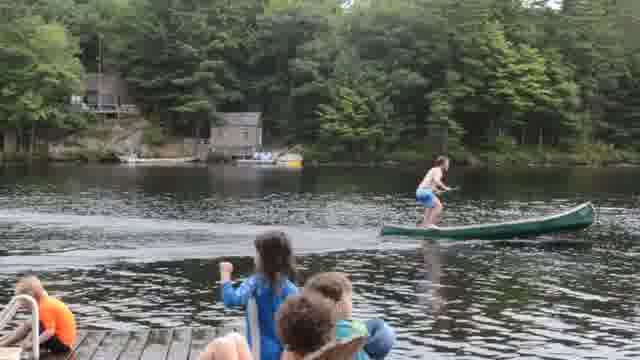}};
\node at (12,0) {\includegraphics[width=0.2\textwidth,trim={10cm 0 0 0},clip]{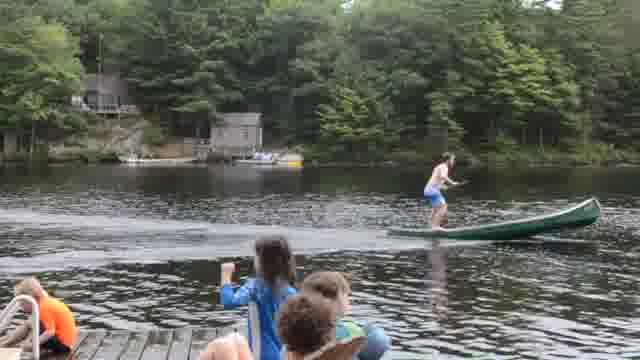}};
\node at (0,-4) {\includegraphics[width=0.2\textwidth,trim={10cm 0 0 0},clip]{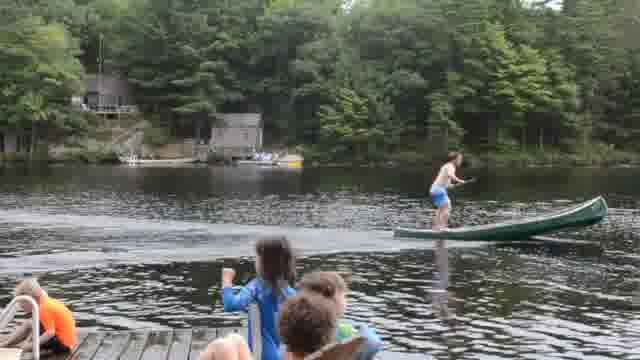}};
\node at (4,-4) {\includegraphics[width=0.2\textwidth,trim={10cm 0 0 0},clip]{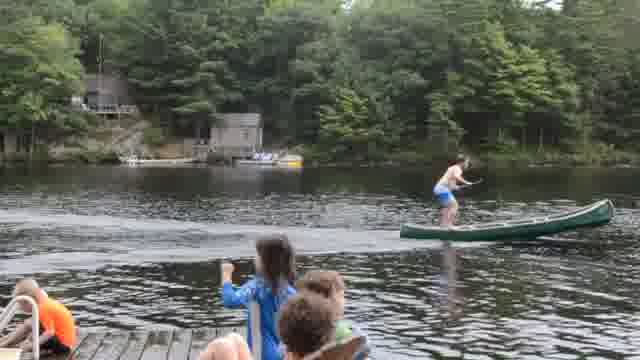}};
\node at (8,-4) {\includegraphics[width=0.2\textwidth,trim={10cm 0 0 0},clip]{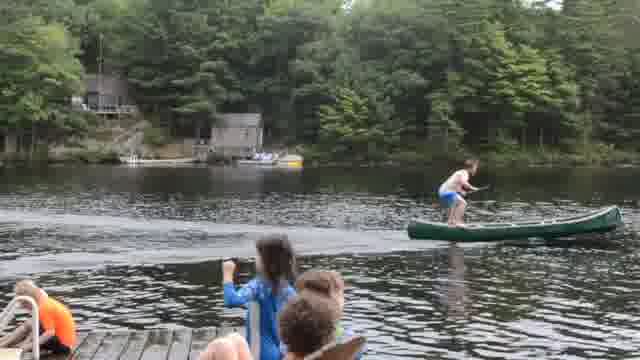}};
\node at (12,-4) {\includegraphics[width=0.2\textwidth,trim={10cm 0 0 0},clip]{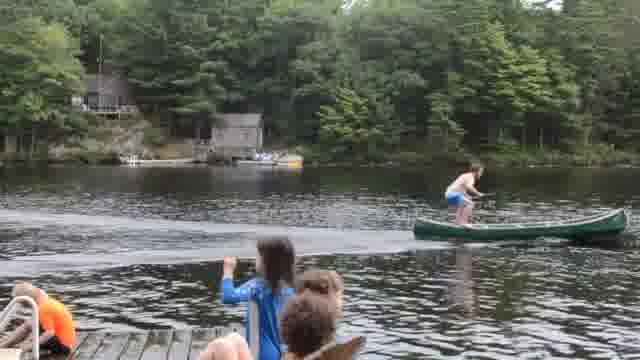}};
\end{tikzpicture}
\caption{Photographs of Jerome A. Neufeld gunwale bobbing on his canoe of dimensions $L,W,D=4.7,0.94,0.15\un{m}$,  taken at intervals of 0.1 s. \label{jeromepics}}
\end{figure}

The gunwale bobbing experiments were performed by Jerome A. Neufeld and Miles Neufeld on Muldrew Lake in Ontario, Canada in August 2021. The accelerometer data used in our study was taken by Jerome A. Neufeld using the \textit{Accelerometer} app on an \textit{iPhone 7} whilst bobbing on a canoe of dimensions $L,W,D=4.7,0.94,0.15\un{m}$. The photographs in Figs. \ref{pitch}a and \ref{thewaves}e of the main text corresponded to gunwale bobbing performed by Miles Neufeld on a paddle board of dimensions $L,W,D=3.05,0.76,0.10\un{m}$. Photographs of Jerome A. Neufeld on his canoe are illustrated here in Fig. \ref{jeromepics} and a video is uploaded as a separate file.

\begin{figure}
\centering
\begin{tikzpicture}[scale=1]
\node at (0,0) {\includegraphics[width=0.9\textwidth]{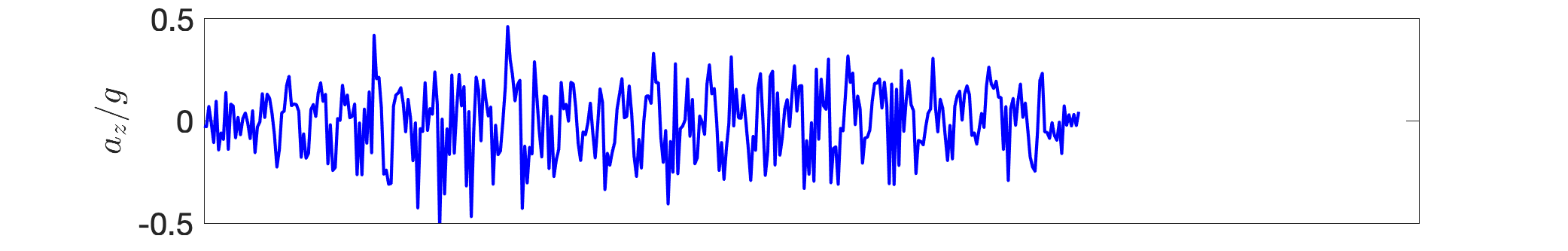}};
\node at (0,-2.5) {\includegraphics[width=0.9\textwidth]{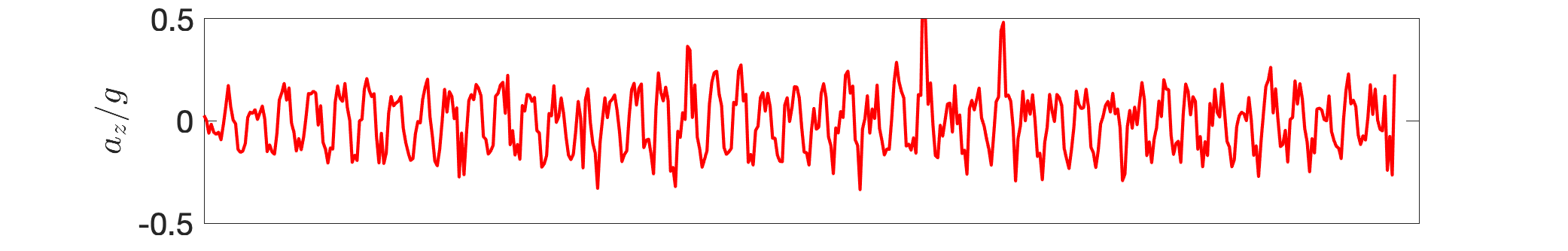}};
\node at (0,-5) {\includegraphics[width=0.9\textwidth]{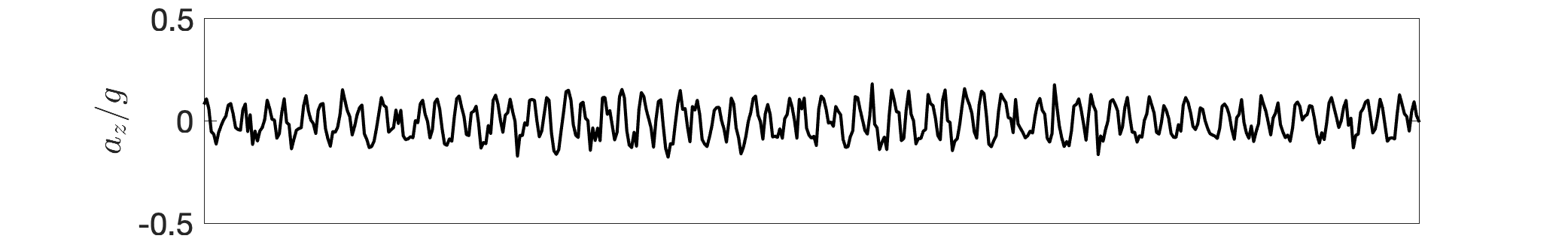}};
\node at (0,-7.5) {\includegraphics[width=0.9\textwidth]{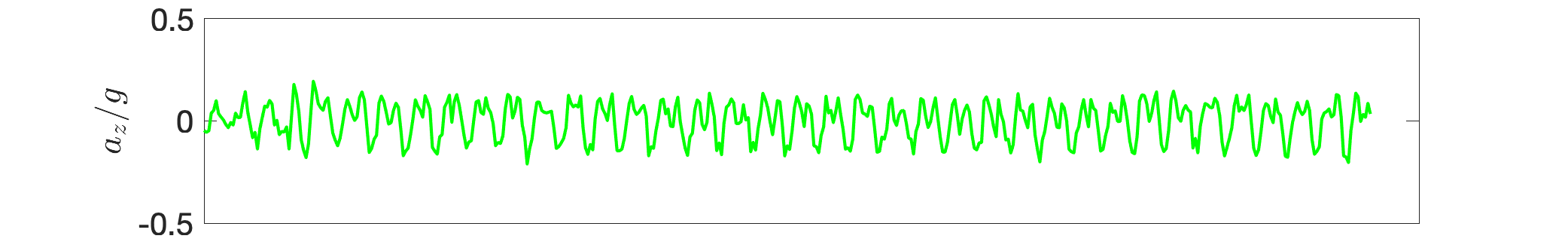}};
\node at (0,-10.3) {\includegraphics[width=0.9\textwidth]{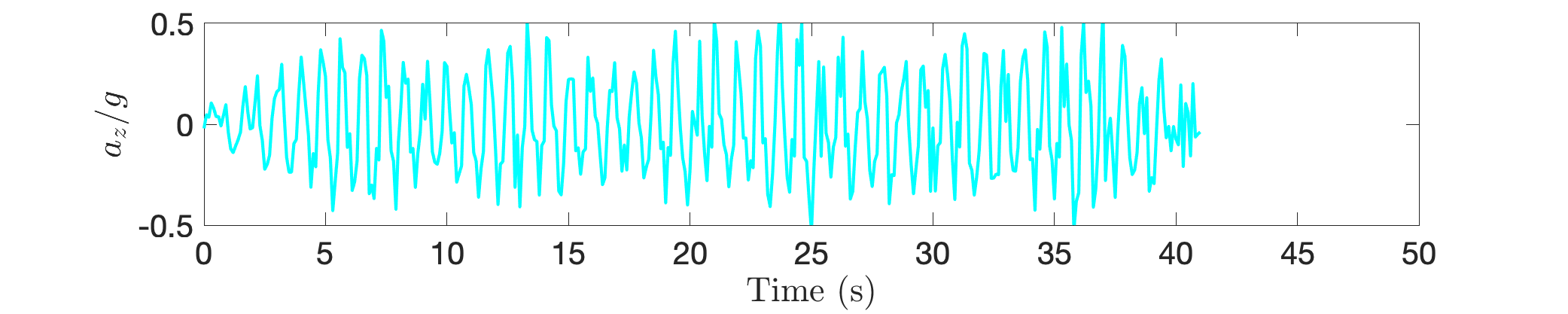}};
\end{tikzpicture}
\caption{Experimental data for the vertical acceleration (normalised by 
$g$) measured using an accelerometer over five separate trials. \label{expdata}}
\end{figure}

Data for the vertical acceleration during gunwale bobbing over five different trials are plotted in Fig. \ref{expdata}. Aside from the trials corresponding to the displayed accelerometer data, a total eight further gunwale bobbing trials were performed by Jerome A. Neufeld over a measured distance of $24.73$ m between two jetties on the lake to measure the speed of the canoe. The times to complete this journey were $T=21.5,21.1,24.66,32.98,29.15,47.53,21.41,20.4$ s. Hence, the average speed plus or minus one standard deviation was $U=0.97\pm0.24$ m/s.


\section{Modelling profile drag}

In this section we briefly describe how we model profile drag, which is given by \eqref{Rskin} in the manuscript.
As described by \cite{S_boucher2018thin}, there is a significant contribution to the drag on a canoe from viscous friction at the wetted surface, and from the form drag due to vortex shedding. The skin and form drag are summed together and modelled with a combined profile drag term $R_d$, given in terms of the dimensionless wetted surface area $\mathcal{S}$ and a drag coefficient $C_d$ (see \eqref{Rskin} in the manuscript).
Following \cite{S_boucher2018thin}, the drag coefficient is approximated by the empirical relationship
\beq
C_d=C_f(1+2/\alpha+60/\alpha^4),\label{dragapprox}
\eeq
where $C_f$ is the skin friction coefficient for a flat plate \cite{S_hoerner1965practical}. This varies weakly with the Reynolds number $\mathrm{Re}=UL/\nu$, where $\nu$ is the kinematic viscosity, and is approximated for turbulent flows \cite{S_hadler1958coefficients} as
\beq
C_f={0.075}(\log\mathrm{Re}-2)^{-2}.
\eeq
For example, a $4.7\un{m}$ canoe cruising at 1 m/s in water corresponds to a Reynolds number of $4.7\times10^6$, producing a skin friction value $C_f=3.4\times 10^{-3}$. 

\end{document}